\documentclass[reprint,showpacs,preprintnumbers,pre,superscriptaddress]{revtex4-1}
\usepackage[T1]{fontenc}
\usepackage[utf8]{inputenc}
\usepackage{bm}
\usepackage{amsmath}
\usepackage{amssymb}
\usepackage{graphicx}

\makeatletter
\usepackage{amsmath, amssymb}   % 提供 \operatorname 和 AMS 功能
\usepackage{bm}
\usepackage{amsfonts}
\usepackage{color}
\usepackage[ruled,linesnumbered]{algorithm2e}

\SetAlgoCaptionSeparator{.}

\SetNlSty{\normalfont\small}{}{:}

\SetKwInput{KwIn}{Input}

\DontPrintSemicolon

\SetAlCapNameFnt{\normalfont}
\SetAlCapFnt{\normalfont}
\newcommand{\avg}[1]{\langle #1 \rangle}
\newcommand{\dd}{\mathrm{d}}

\makeatother

\begin{document}
\title{Shortcuts to Parameter Sweeps}
\author{Chi Xiang}
\affiliation{School of Systems Science, Beijing Normal University, Beijing 100875,
China}
\author{Guodong Cheng}
\affiliation{School of Physics and Astronomy, Beijing Normal University, Beijing
100875, China}
\author{Geng Li}
\email{gengli@bnu.edu.cn}

\affiliation{School of Systems Science, Beijing Normal University, Beijing 100875,
China}
\begin{abstract}
Efficient evaluation of stationary parametric sensitivities over broad
parameter ranges is important for identifying influential training
data, fitting force fields, and predicting material responses, but
standard pointwise approaches require repeated relaxation and sampling.
Here we introduce Shortcuts to Parameter Sweeps (STPS), an engineered
control strategy that uses an auxiliary control to transport the probability
density along a prescribed family of instantaneous stationary states
during a finite-time parameter sweep. This enables the continuous
response curve over the full parameter interval to be estimated from
a single controlled sweep using covariance-based response relations.
STPS applies to both equilibrium and nonequilibrium steady-state systems,
including those with unknown stationary distributions, and can be
implemented directly using stationary samples in high-dimensional
settings. Numerical tests on single-particle and interacting many-body
systems show that STPS yields response curves in close agreement with
reference results. These findings establish STPS as an efficient,
sample-based framework for continuous sensitivity analysis in stochastic
simulations.
\end{abstract}
\maketitle
\emph{Introduction.}--Stationary parametric responses quantify how
steady-state expectations vary with externally controlled parameters.
Accurate characterization of these sensitivities is useful for identifying
influential training examples in machine-learning models \citep{koh2017},
calibrating force-field parameters against reference data \citep{Wang2014},
and predicting nonequilibrium material response under changes in interactions
or loading conditions \citep{Huang2022}. Despite their importance,
efficiently estimating responses over a continuous range of parameter
values remains challenging. Finite-difference methods approximate
the local response at a selected parameter value using simulations
at nearby parameter values \citep{Zazanis_1993,Anderson_2012}. Their
accuracy depends critically on the perturbation size, which must balance
finite-difference truncation error against the amplification of sampling
noise \citep{Gill_1983,More2012}. Repeating this procedure throughout
a parameter interval can therefore be computationally expensive, as
long equilibration and sampling runs are required at each parameter
value, particularly for slowly relaxing systems. Trajectory-weight
methods instead estimate parameter sensitivities by differentiating
the probability measure over trajectories, thereby avoiding simulations
at perturbed parameter values. This principle underlies likelihood-ratio
estimators for a broad class of stochastic dynamics \citep{Glynn_1990,Plyasunov2007,Arampatzis_2016}.
It also underlies Malliavin-weight sampling, which is grounded in
Malliavin calculus \citep{Bell2006,Nualart2006}, was developed for
sensitivity estimation in Brownian-dynamics simulations \citep{Warren2012},
and was subsequently formulated within a more general framework for
stochastic simulation algorithms \citep{Warren2014}. These estimators
provide local sensitivities at the simulated parameter value, but
their variance can grow with trajectory length when uncentered weights
are accumulated. More recently, a score-shifted stochastic differential
equation approach has enabled path reweighting estimates to diffusion
perturbations by mapping them to effective drift perturbations \citep{Klinger2025}.
Score-based generative models have also been used to estimate transient
linear responses through the generalized fluctuation-dissipation theorem
\citep{Giorgini_2024}. Despite these advances, efficiently computing
continuous response curves over broad parameter ranges within a single
finite-time protocol remains challenging for both equilibrium and
nonequilibrium steady-state systems.

Recently, shortcut methods have been extensively developed and applied
in quantum systems \citep{Demirplak2003,Berry_2009,Chen_2010,Torrontegui2013,GueryOdelin2019}
and increasingly extended to biophysical settings \citep{Iram2021,Ilker2022},
offering a promising route to overcoming this difficulty. Broadly,
shortcuts are finite-time controls that modify the bare dynamics with
an auxiliary control to realize the target state that would otherwise
require a slow adiabatic or quasi-static process \citep{GueryOdelin_2023}.
In quantum systems, these methods are known as shortcuts to adiabaticity
\citep{Torrontegui2013,GueryOdelin2019}. In stochastic systems, related
techniques such as counterdiabatic driving \citep{Iram2021,Ilker2022},
engineered swift equilibration \citep{Martinez2016}, and shortcuts
to isothermality \citep{Li2017} similarly accelerate state transitions
and enable the target distribution to be reached within a finite time.

Here, we apply the shortcut idea for estimating response functions
during continuous parameter sweeps, with the aim of efficiently estimating
response over a wide parameter range from a single engineered sweep.
For equilibrium systems, the stationary distribution is known as the
Boltzmann distribution, and the response is computed using the standard
equilibrium covariance identity \citep{Arsham_1989,Zhu1993,Zheng_2026},
with the instantaneous stationary distribution realized by using the
shortcut method for a wide range of parameter changes. For nonequilibrium
steady-state systems, the stationary distribution is generally not
available in closed form \citep{Seifert2012,BaiesiMaes2013}. We first
use a graph neural network to learn the instantaneous state-space
score, which characterizes the local structure of the distribution,
and then use the learned score to construct the auxiliary control
that guides the system through the instantaneous stationary distributions
during finite-time control processes. The covariance-based response
relation remains valid, but its direct computation is generally difficult
for nonequilibrium steady-state systems. We reformulate this relation
in terms of the learned auxiliary control, enabling response functions
to be computed directly from nonequilibrium steady-state samples.

\begin{figure}[!tp]
\centering{}\includegraphics[width=8.5cm]{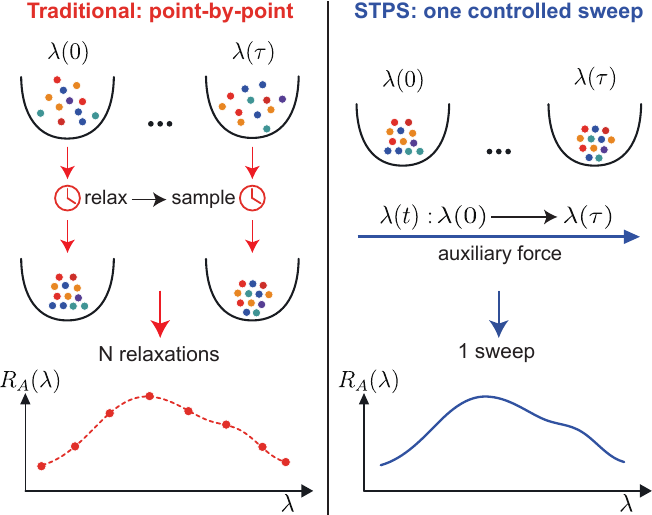} \caption{Point-by-Point Relaxation versus Single-Sweep Continuous Response
Estimation. Left: In conventional approaches, the system is simulated
independently at a sequence of discrete parameter values $\lambda_{i}$.
At each parameter value, the system must first relax to the corresponding
stationary state before the observable $R_{A}(\lambda_{i})$ is sampled.
The response curve is subsequently reconstructed from these discrete
measurements. Right: In STPS, the control parameter $\lambda(t)$
is continuously swept from $\lambda(0)$ to $\lambda(\tau)$, while
an auxiliary control ensures that the system evolves according to
the instantaneous stationary-state distribution throughout the sweep.
Consequently, STPS estimates the continuous response curve $R_{A}(\lambda)$
over the entire parameter interval from a single finite-time sweep,
without requiring repeated relaxation and independent sampling at
individual parameter values.}
\end{figure}

We validate Shortcuts to Parameter Sweeps (STPS) using equilibrium
and nonequilibrium steady-state models. The equilibrium examples include
a single particle confined in a translated harmonic trap and a many-body
system with pairwise Gaussian-core interactions, while the nonequilibrium
steady-state example is a rotating many-body system. Across these
examples, STPS produces response curves that agree well with the theoretical
reference curves over a broad parameter range.

\emph{Universal stationary response.}--Let $\lambda$ denote a scalar
externally controlled parameter that determines the instantaneous
stationary state of the system. During a finite-time protocol, the
parameter becomes time dependent, $\lambda=\lambda(t)$. Let $\rho_{{\rm \mathrm{st}}}(\bm{x},\lambda)$
be a normalized stationary distribution on $\Omega\subseteq\mathbb{R}^{n}$,
and $A(\bm{x},\lambda)$ an observable. The stationary-state average
of an observable $A$ is defined as $\avg{A}_{\lambda}=\int_{\Omega}A(\bm{x},\lambda)\,\rho_{{\rm \mathrm{st}}}(\bm{x},\lambda)\,\dd^{n}x.$
We introduce the generalized potential $\Phi(\bm{x},\lambda)$ through
the stationary-state distribution, $\rho_{{\rm \mathrm{st}}}(\bm{x},\lambda)=e^{-\Phi(\bm{\bm{x}},\lambda)}$
with $\Phi(\bm{x},\lambda)\equiv-\ln\rho_{{\rm \mathrm{st}}}(\bm{x},\lambda)$.
Differentiating the stationary average with respect to $\lambda$
yields \citep{Arsham_1989,Zheng_2026}
\begin{equation}
R_{A}(\lambda)\equiv\partial_{\lambda}\langle A\rangle_{\lambda}=\left\langle \partial_{\lambda}A\right\rangle _{\lambda}-\operatorname{Cov}_{\lambda}\left(A,\partial_{\lambda}\Phi\right),\label{eq:stationary-response}
\end{equation}
where $\operatorname{Cov}_{\lambda}(A,B)\equiv\langle AB\rangle_{\lambda}-\langle A\rangle_{\lambda}\langle B\rangle_{\lambda}.$
This identity holds exactly for any normalized stationary density
$\rho_{{\rm \mathrm{st}}}(\bm{x},\lambda)$ that is sufficiently smooth
with respect to $\lambda$. At equilibrium, the stationary density
takes the Boltzmann form $\rho_{{\rm \mathrm{st}}}(\bm{x},\lambda)\propto\exp[-\beta H_{\mathrm{o}}(\bm{x},\lambda)]$,
where $H_{\mathrm{o}}(\bm{x},\lambda)$ denotes the Hamiltonian of
the system. The above relation then reduces to $R_{A}(\lambda)=\left\langle \partial_{\lambda}A\right\rangle _{\lambda}-\beta\operatorname{Cov}_{\lambda}\left(A,\partial_{\lambda}H_{\mathrm{o}}\right)$,
which is the equilibrium response relation discussed in \citep{Zhu1993}.
Importantly, Eq.$~\eqref{eq:stationary-response}$ is expressed entirely
in terms of averages over the stationary ensemble at a fixed value
of $\lambda$. Therefore, to evaluate the response curve while sweeping
the parameter according to $\lambda(t)$, the samples collected at
each time $t$ must be distributed according to the corresponding
instantaneous stationary density, $p(\bm{x},t)=\rho_{{\rm \mathrm{st}}}(\bm{x},\lambda(t))$.
Under the bare dynamics, however, varying $\lambda$ at a finite rate
generally drives the system away from the instantaneous stationary
state, resulting in a lag \citep{Vaikuntanathan2008}, $p(\bm{x},t)\neq\rho_{{\rm \mathrm{st}}}(\bm{x},\lambda(t)).$
Consequently, directly applying Eq.$~\eqref{eq:stationary-response}$
to samples collected during an uncontrolled finite-time sweep generally
yields a biased estimate of the stationary response.

\emph{Stationary transport.--}To eliminate this lag, we introduce
an auxiliary control $\bm{f}$ and consider the controlled dynamics
\begin{equation}
\dd\bm{X}_{t}=\left[\bm{b}_{\mathrm{o}}(\bm{X}_{t},\lambda)-\bm{f}(\bm{X}_{t},t)\right]\dd t+\sqrt{2D}\,\dd\bm{W}_{t},\label{eq:controlled-dynamics}
\end{equation}
where $\bm{b}_{\mathrm{o}}$ denotes the bare drift, whose stationary
density at fixed $\lambda$ is $\rho_{{\rm \mathrm{st}}}(\bm{x},\lambda)$.
Here, $D$ is the diffusion coefficient that determines the strength
of the stochastic fluctuations, and $\bm{W}_{t}$ is a standard Wiener
process. Because the required correction vanishes in the quasi-static
limit and reverses sign upon reversal of the protocol, we write the
auxiliary control as $\bm{f}(\bm{X}_{t},t)=\dot{\lambda}\,\bm{u}(\bm{X}_{t},\lambda)$,
where $\bm{u}$ is an escort field that transports the probability
density along the family of instantaneous stationary states. The Supplemental
Material \citep{sm} verifies that this construction yields exact
stationary tracking under suitable regularity and boundary conditions.

We then require the probability density generated by Eq.$~\eqref{eq:controlled-dynamics}$
to remain on this stationary manifold $p(\bm{x},t)=\rho_{{\rm \mathrm{st}}}(\bm{x},\lambda(t))$
throughout the protocol$.$ Substituting this condition into the corresponding
Fokker--Planck equation of Eq.$~\eqref{eq:controlled-dynamics}$,
and using the fact that $\rho_{{\rm \mathrm{st}}}(\bm{x},\lambda)$
is stationary under the bare drift $\bm{b}_{\mathrm{o}}$ at fixed
$\lambda$, yields the transport equation
\begin{equation}
\partial_{\lambda}\rho_{{\rm st}}=\nabla\!\cdot\left(\bm{u}\rho_{{\rm \mathrm{st}}}\right).\label{eq:transport}
\end{equation}
Accordingly, an escort field $\bm{u}$ satisfying this equation transports
the ensemble along the stationary manifold and eliminates finite-rate
distributional lag. This enables continuous evaluation of response
functions throughout the sweep. Previous studies have used this transport
equation to realize finite-time state transformations \citep{Li2017,Patra2017},
whereas here we firstly connect it to the calculation of response
functions. We refer to the resulting framework, which combines Eqs.$~\eqref{eq:stationary-response}$
and$~\eqref{eq:transport}$ to evaluate response functions during
a single controlled parameter sweep, as Shortcuts to Parameter Sweeps.

\emph{Response of unknown stationary densities.--}For a system under
persistent nonequilibrium driving, such as coupling to multiple reservoirs
or the action of nonconservative forces, suitable ergodicity and confinement
conditions typically lead to relaxation toward a nonequilibrium steady
state. Such a state generally violates detailed balance and may sustain
a nonzero stationary probability current \citep{Seifert2012}. The
stationary response considered here, however, is determined by the
parameterized stationary density $\rho_{\mathrm{st}}(\bm{x},\lambda)$,
rather than by the current itself. The main practical difficulty is
that $\rho_{\mathrm{st}}(\bm{x},\lambda)$ is rarely available in
closed form \citep{Seifert2012,BaiesiMaes2013}. Obtaining it generally
requires solving the corresponding stationary Fokker-Planck equation,
which is often analytically difficult and whose direct numerical discretization
suffers from the curse of dimensionality \citep{Sun2014}. Consequently,
the generalized potential $\Phi(\bm{x},\lambda)=-\ln\rho_{\mathrm{st}}(\bm{x},\lambda)$
is typically unknown as well \citep{Seifert2012,BaiesiMaes2013}.
We therefore consider a sample-access setting in which stationary
configurations can be generated by stochastic simulations at selected
values of $\lambda$. This motivates a representation of the stationary
density that can be inferred directly from samples, without explicitly
reconstructing the stationary density over the high-dimensional configuration
space \citep{GutmannCorander2016,LiTurner2018,ZhouShiZhu2020}. 

A natural such representation is the stationary score \citep{Hyvarinen2005},
defined as $\bm{s}(\bm{x},\lambda)\equiv\nabla\ln\rho_{\mathrm{st}}(\bm{x},\lambda)$,
where $\nabla$ denotes the gradient with respect to $\bm{x}$. Score-matching
methods allow $\bm{s}$ to be estimated directly from stationary samples,
without evaluating either the density or its normalization constant
\citep{LiTurner2018,ZhouShiZhu2020}. In addition to characterizing
the local structure of the stationary density, the score defines an
effective reversible drift $\bm{b}_{\mathrm{eff}}(\bm{x},\lambda)=D\bm{s}(\bm{x},\lambda)$.
The associated probability current is $\bm{j}_{\mathrm{eff}}=\bm{b}_{\mathrm{eff}}\rho_{\mathrm{st}}-D\nabla\rho_{\mathrm{st}}$,
which vanishes identically because $\bm{s}\rho_{\mathrm{st}}=\nabla\rho_{\mathrm{st}}$.
The resulting effective dynamics therefore has the same stationary
density as the original nonequilibrium system, although it does not
reproduce its physical drift or stationary current. Since the controlled-dynamics
framework only requires a bare drift whose stationary density is $\rho_{\mathrm{st}}$,
we may set $\bm{b}_{\mathrm{o}}=\bm{b}_{\mathrm{eff}}$ in Eq.$~\eqref{eq:controlled-dynamics}$.
This construction enables stationary transport using only sampled
configurations, without explicitly determining either $\rho_{\mathrm{st}}$
or the underlying physical dynamics.

The remaining ingredient required by the stationary-response identity
in Eq.$~\eqref{eq:stationary-response}$ is the parameter derivative
of the logarithmic stationary density, $\partial_{\lambda}\ln\rho_{{\rm \mathrm{st}}}(\bm{x},\lambda)$.
Directly evaluating this quantity is generally challenging because
the stationary density itself is not known explicitly. In particular,
reconstructing stationary densities at nearby parameter values is
computationally demanding in high-dimensional systems because multivariate
density estimation suffers from sample sparsity \citep{Scott_1991}.
Numerically differentiating these noisy density estimates further
amplifies sampling error and introduces a finite-difference step-size
tradeoff \citep{More2012}. We circumvent this difficulty by exploiting
the transport equation introduced above in Eq.$~\eqref{eq:transport}$.
Dividing it by $\rho_{{\rm \mathrm{st}}}(\bm{x},\lambda)$ yields
\begin{equation}
G(\bm{x},\lambda)\equiv\partial_{\lambda}\ln\rho_{{\rm \mathrm{st}}}(\bm{x},\lambda)=\nabla\cdot\bm{u}(\bm{x},\lambda)+\bm{u}(\bm{x},\lambda)\cdot\bm{s}(\bm{x},\lambda).\label{eq:G-identity}
\end{equation}
Thus, the combination of the stationary score $\bm{s}$ and the escort
field $\bm{u}$ provides $\partial_{\lambda}\ln\rho_{{\rm \mathrm{st}}}$
directly, without reconstructing the normalized density. Since $\partial_{\lambda}\Phi=-G$,
substitution into Eq.$~\eqref{eq:stationary-response}$ gives 
\begin{equation}
R_{A}(\lambda)=\left\langle \partial_{\lambda}A\right\rangle _{\lambda}+\operatorname{Cov}_{\lambda}\left(A,G\right).\label{eq:continuous-response}
\end{equation}
If the stationary density $\rho_{{\rm \mathrm{st}}}(\bm{x},\lambda)$
is explicitly known, the response can be computed using Eq.$~\eqref{eq:stationary-response}$.
When the stationary density is unavailable, we instead evaluate the
response using Eq.$~\eqref{eq:continuous-response}$, which provides
an operational formula that does not require explicit knowledge of
the normalized stationary density. Once the score and an escort field
satisfying Eq.$~\eqref{eq:transport}$ have been obtained, the responses
of different observables can be evaluated along the same controlled
parameter protocol, without requiring exact knowledge of either $\rho_{{\rm \mathrm{st}}}$
or the physical drift that generated it. For interacting many-body
systems, we represent the score $\bm{s}$ and the escort field $\bm{u}$
using graph neural networks. Specifically, $\bm{s}_{\phi}(\bm{x},\lambda)$
and $\bm{u}_{\theta}(\bm{x},\lambda)$ denote neural-network parameterizations
of the score and the escort field, respectively, with $\phi$ and
$\theta$ representing the corresponding trainable parameters. After
offline training, the quantity entering Eq.$~\eqref{eq:continuous-response}$
is evaluated as $G_{\theta,\phi}(\bm{x},\lambda)=\nabla\cdot\bm{u}_{\theta}(\bm{x},\lambda)+\bm{u}_{\theta}(\bm{x},\lambda)\cdot\bm{s}_{\phi}(\bm{x},\lambda).$
The network architectures, learning objectives, transport regularization,
and training procedures are described in the Supplemental Material
\citep{sm}.

\begin{figure*}[!t]
\centering{}\includegraphics[width=1\textwidth]{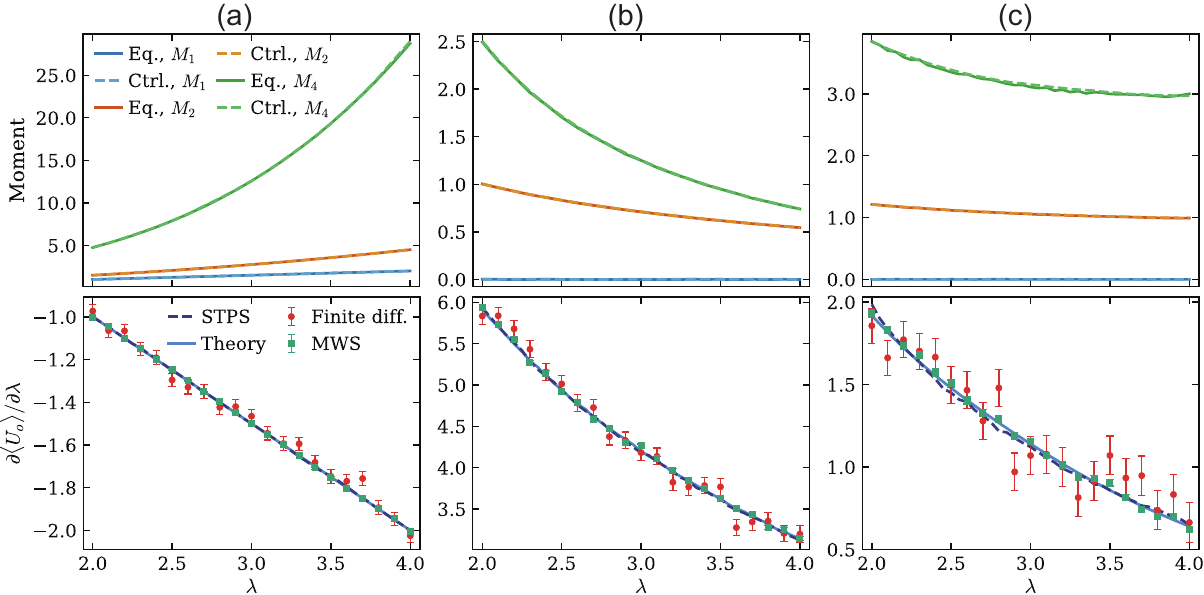} \caption{Numerical validation of STPS for continuous response estimation. (a)
Translated harmonic trap, $U_{\mathrm{o}}=kx^{2}/2-\lambda x$, with
$k=2$. (b) Interacting many-body system, $U_{\mathrm{o}}=a\sum_{i<j}e^{-r_{ij}^{2}/(2\sigma^{2})}+\frac{\lambda}{2}\sum_{i}\lVert\bm{r}_{i}\rVert^{2}$
with $N=10$, $\sigma=1$, and $a=1$. (c) Rotating interacting many-body
system, $U_{\mathrm{o}}(\bm{r};\lambda)=a\sum_{i<j}\exp\left(-r_{ij}^{2}/2\sigma^{2}\right)+\frac{1}{2}\sum_{i}\left(bx_{i}^{2}+\lambda y_{i}^{2}\right)$
subject to the nonconservative rotational force $\omega\bm{R}\bm{r}_{i}$,
with $N=8$, $\sigma=1$, $a=1$, $b=1$, and $\omega=1.5$. In all
simulations, the observable is $A=U_{\mathrm{o}}$, $k_{{\rm B}}T=1$,
the control parameter is varied from $\lambda=2$ to $4$ over a protocol
duration $\tau=2$, and the integration time step is $10^{-3}$. Averages
are evaluated over $10^{5}$ independent trajectories. Throughout,
$M_{k}$ denotes the $k$-th raw coordinate moment, averaged over
all particles and Cartesian components: $M_{k}=\frac{1}{dN}\sum_{i=1}^{N}\sum_{\alpha=1}^{d}\left\langle r_{i,\alpha}^{k}\right\rangle $.
For a single one-dimensional harmonic oscillator, $N=d=1$ and $r_{1,1}=x$,
so this definition reduces to $M_{k}=\langle x^{k}\rangle$. \label{fig:eqbenchmarks}}
\end{figure*}

\emph{Numerical tests.--}We first test STPS in two equilibrium systems,
for which the stationary distributions are explicitly known and reference
responses can be obtained analytically or through direct numerical
calculations. In these equilibrium simulations, an auxiliary control
of the form $\dot{\lambda}\bm{u}(\bm{X}_{t},\lambda)$ is applied
to keep the evolving ensemble in the instantaneous equilibrium density
throughout a finite-time parameter sweep.

The first example is a particle confined in a translated harmonic
trap, $U_{\mathrm{o}}(x,\lambda)=kx^{2}/2-\lambda x,$ where varying
$\lambda$ continuously shifts the trap center without changing its
shape. In this case, Eq.$~\eqref{eq:transport}$ can be solved analytically,
giving the auxiliary force $f=-\dot{\lambda}/k$ \citep{Li2017}.
This system therefore provides a simple benchmark in which the exact
shortcut control is known.

The second equilibrium example is a two-dimensional interacting many-body
system with potential $U_{\mathrm{o}}=a\sum_{i<j}e^{-r_{ij}^{2}/(2\sigma^{2})}+\frac{\lambda}{2}\sum_{i}\lVert\bm{r}_{i}\rVert^{2}$
with $r_{ij}=\lVert\bm{r}_{i}-\bm{r}_{j}\rVert$. The first term is
a Gaussian-core pair interaction, where $a>0$ sets the repulsive
strength and $\sigma$ determines the interaction range. Because the
potential remains finite at complete overlap, it provides a standard
soft-core description of penetrable particles \citep{Stillinger1976,Prestipino2011,Cheng2026}.
The second term is an isotropic harmonic trap that confines the particles
near the origin. During the protocol, $\lambda$ is varied to change
the confinement strength. Unlike the translated harmonic trap, the
escort field for this high-dimensional system is not available in
closed form. Instead, $\bm{u}_{\theta}(\bm{x}$,$\lambda)$ is represented
and learned using a graph neural network. Further details on the network
architecture and training procedure are provided in the Supplemental
Material \citep{sm}.

We then test STPS in a nonequilibrium steady-state system, for which
neither the stationary distribution nor the corresponding shortcut
control is known explicitly. Specifically, we consider a two-dimensional
rotating interacting many-body system with conservative potential
$U_{\mathrm{o}}(\bm{r};\lambda)=a\sum_{i<j}\exp\left(-r_{ij}^{2}/2\sigma^{2}\right)+\frac{1}{2}\sum_{i}\left(bx_{i}^{2}+\lambda y_{i}^{2}\right),$
and a nonconservative rotational force $\omega\bm{R}\bm{r}_{i}=\omega(-y_{i},x_{i})$,
where $\bm{R}$ is the two-dimensional skew-symmetric rotation matrix
corresponding to a $90^{\circ}$ counterclockwise rotation in the
plane. Here, $\lambda$ controls the confinement strength along the
$y$ direction and thus tunes the anisotropy of the nonequilibrium
stationary distribution. Because this stationary distribution is unavailable
in closed form, both the stationary score $\bm{s}_{\phi}(\bm{x},\lambda)$
and the shortcut field $\bm{u}_{\theta}(\bm{x},\lambda)$ are learned
from steady-state samples using two separate graph neural networks.

As shown in Fig.$~\ref{fig:eqbenchmarks}$, we benchmark STPS in three
settings of increasing complexity: a translated harmonic trap, an
equilibrium interacting many-body system, and a rotating nonequilibrium
many-body system. During the same finite-time sweep from $\lambda=2$
to $4$, the STPS estimate of the stationary response, $R_{U_{\mathrm{o}}}(\lambda)=\partial_{\lambda}\langle U_{\mathrm{o}}\rangle_{\lambda}$,
agrees throughout the interval with the corresponding theoretical
results. For the harmonic trap, the analytically derived shortcut
reproduces the exact theoretical curve, directly validating both the
controlled transport and the response estimator. For the equilibrium
many-body system, the graph-neural-network-learned escort field remains
accurate in a high-dimensional interacting configuration space, despite
the absence of a closed-form shortcut. The rotating system provides
a more stringent test: detailed balance is broken, and neither the
stationary density nor the exact shortcut is available. Nevertheless,
the jointly learned score and escort fields recover the reference
response without reconstructing the normalized density or identifying
the original physical drift. The lower panels provide post-training
distributional consistency diagnostics: the controlled moments $M_{1}$,
$M_{2}$, and $M_{4}$ closely track their instantaneous stationary
counterparts throughout the sweep. Their simultaneous agreement is
consistent with the controlled ensemble remaining close to the target
stationary path over the full protocol. Taken together, these results
show that STPS suppresses finite-rate distributional lag and accurately
reconstructs continuous stationary-response curves from a single finite-time
sweep in both equilibrium and nonequilibrium many-body systems.

\emph{Discussion.}--STPS recasts stationary-response estimation over
a parameter interval as a distribution-transport problem. At equilibrium,
it combines the covariance identity underlying linear response with
escort transport. STPS exploits the entire transported stationary
path, enabling a continuous response curve over a broad parameter
interval to be estimated from an ensemble of realizations evolving
under a single finite-time sweep protocol. For a nonequilibrium steady
state with an unknown stationary density, the learned score and escort
field jointly determine $G=\partial_{\lambda}\ln\rho_{{\rm \mathrm{st}}}$,
so that the response can be evaluated without reconstructing the normalized
stationary density or identifying the underlying physical drift. STPS
thus provides a sample-based approach to stationary response estimation
that complements generalized fluctuation-response theories for nonequilibrium
steady states \citep{Prost2009,Seifert2010} and draws on score matching
to estimate log-density gradients without evaluating normalization
constants \citep{Hyvarinen2005}.

The formal construction is exact when the stationary score and escort
field are exact, whereas a learned implementation introduces several
coupled errors: finite coverage of the training distributions, score
error, residual violation of the transport equation, numerical error
in $\nabla\!\cdot\bm{u}$, and covariance sampling noise. These effects
may become pronounced in metastable regimes or near sharp stationary
crossovers, where mixing is slow and neighboring distributions have
weak overlap. Quantitative error bounds, cost-accuracy scaling, and
diagnostics based on transport residuals therefore remain important
next steps.

Several extensions follow naturally. For a vector-valued control parameter
($\boldsymbol{\lambda}=(\lambda_{1},\ldots,\lambda_{m})$), the present
transport framework could be generalized by learning a distinct escort
field $\bm{u}_{\alpha}$ for each parameter direction $\lambda_{\alpha}$,
in structural analogy with the direction-dependent adiabatic gauge
potentials defined over parameter space \citep{Kolodrubetz2017}.
Larger many-body applications should benefit from architectures that
enforce permutation and Euclidean symmetries \citep{Satorras2021,Batzner2022}.
Finally, systematic comparisons with pointwise relaxation, likelihood-ratio
estimators \citep{Glynn_1990,Plyasunov2007}, and Malliavin-weight
sampling \citep{Warren2012,Warren2014} should identify the regimes
in which the offline learning and control cost of STPS is offset by
reusing one sweep to evaluate many parameter values and observables.

\emph{Acknowledgements.}--We are grateful to Professors Zhanchun
Tu and Zhiyue Lu for their valuable suggestions and insightful comments
on this research. This work is supported by the National Natural Science
Foundation of China (NSFC) (Grant No. 12405031). G. L. also acknowledges
the financial support from the Zhongying Young Scholars Program of
Beijing Normal University.

\bibliographystyle{apsrev4-1}
\bibliography{refs_main}

\end{document}

\title{Supplementary materials: Shortcuts to Parameter Sweeps}
\author{Chi Xiang}
\affiliation{School of Systems Science, Beijing Normal University, Beijing 100875,
China}
\author{Guodong Cheng}
\affiliation{School of Physics and Astronomy, Beijing Normal University, Beijing
100875, China}
\author{Geng Li}
\email{gengli@bnu.edu.cn}

\affiliation{School of Systems Science, Beijing Normal University, Beijing 100875,
China}

\maketitle
The supplementary materials are devoted to provide detailed derivations
in the main context.

\tableofcontents{}

\section{Stationary response and exact transport}

\subsection{Stationary response identity}

Let $\lambda\in\mathcal{I}\subset\mathbb{R}$ be a scalar control
parameter, and let $\bm{x}\in\Omega\subseteq\mathbb{R}^{n}$ denote
the state of the system. At fixed $\lambda$, consider the stochastic
differential equation
\begin{equation}
\dd\bm{X}_{t}=\bm{b}_{\mathrm{o}}(\bm{X}_{t},\lambda)\,\dd t+\sqrt{2D}\,\dd\bm{W}_{t},\label{eq:supp-bare-sde}
\end{equation}
where $\bm{b}_{\mathrm{o}}:\Omega\times\mathcal{I}\to\mathbb{R}^{n}$
is the bare drift, $D>0$ is a constant scalar diffusion coefficient,
and $\bm{W}_{t}$ is a standard $n$-dimensional Wiener process. We
write $\nabla\equiv\nabla_{\bm{x}}$. The probability density $p(\bm{x},t)$
obeys
\begin{equation}
\partial_{t}p=\mathcal{L}_{\lambda}^{\dagger}p\equiv-\nabla\cdot\left[\bm{b}_{\mathrm{o}}(\bm{x},\lambda)p\right]+D\nabla^{2}p.\label{eq:supp-bare-fpe}
\end{equation}

For each $\lambda\in\mathcal{I}$, we assume that the stochastic differential
equation and its Fokker-Planck equation are well posed and that the
fixed-$\lambda$ dynamics admit a unique, normalized, and strictly
positive stationary density $\rho_{\mathrm{st}}(\bm{x},\lambda)$
satisfying 
\begin{equation}
\mathcal{L}_{\lambda}^{\dagger}\rho_{\mathrm{st}}(\bm{x},\lambda)=0,\qquad\int_{\Omega}\rho_{\mathrm{st}}(\bm{x},\lambda)\,\dd^{n}x=1.\label{eq:supp-stationary-fpe}
\end{equation}
The state space $\Omega$ is independent of $\lambda$. The stationary
density is assumed to be sufficiently smooth in $\bm{x}$ and continuously
differentiable in $\lambda$. 

Let $A(\bm{x},\lambda)$ be a scalar observable that is continuously
differentiable in $\bm{x}$ and $\lambda$. We assume that $A$, $\partial_{\lambda}A$,
and the products appearing below are integrable under $\rho_{\mathrm{st}}$,
and that differentiation with respect to $\lambda$ may be interchanged
with integration. Its stationary expectation is
\begin{equation}
\langle A\rangle_{\lambda}=\int_{\Omega}A(\bm{x},\lambda)\rho_{\mathrm{st}}(\bm{x},\lambda)\,\dd^{n}x.\label{eq:supp-stationary-average}
\end{equation}
Define the generalized potential $\Phi(\bm{x},\lambda)=-\ln\rho_{\mathrm{st}}(\bm{x},\lambda).$
Then $\partial_{\lambda}\rho_{\mathrm{st}}=-\rho_{\mathrm{st}}\,\partial_{\lambda}\Phi$
with $\left\langle \partial_{\lambda}\Phi\right\rangle _{\lambda}=0$,
where the second identity follows by differentiating the normalization
in Eq.$~\eqref{eq:supp-stationary-fpe}$. Differentiating Eq.$~\eqref{eq:supp-stationary-average}$
therefore gives
\begin{align}
R_{A}(\lambda) & \equiv\partial_{\lambda}\langle A\rangle_{\lambda}\nonumber \\
 & =\left\langle \partial_{\lambda}A\right\rangle _{\lambda}-\left\langle A\,\partial_{\lambda}\Phi\right\rangle _{\lambda}\nonumber \\
 & =\left\langle \partial_{\lambda}A\right\rangle _{\lambda}-\Cov_{\lambda}\!\left(A,\partial_{\lambda}\Phi\right),\label{eq:supp-universal-response}
\end{align}
where $\Cov_{\lambda}(A,B)=\langle AB\rangle_{\lambda}-\langle A\rangle_{\lambda}\langle B\rangle_{\lambda}.$
Under the stated regularity assumptions, Eq.$~\eqref{eq:supp-universal-response}$
holds for any differentiable normalized family of densities and does
not require detailed balance \citep{Zheng2026}. 

At equilibrium,
\begin{equation}
\rho_{\mathrm{st}}(\bm{x},\lambda)=Z(\lambda)^{-1}\exp\!\left[-\beta H_{\mathrm{o}}(\bm{x},\lambda)\right],\label{eq:supp-equilibrium-density}
\end{equation}
where $H_{\mathrm{o}}(\bm{x},\lambda)$ denotes the Hamiltonian of
the system and $Z(\lambda)=\int\exp\!\left[-\beta H_{\mathrm{o}}(\bm{x},\lambda)\right]\dd^{n}x$
is the corresponding partition function. Since $\partial_{\lambda}\ln Z(\lambda)$
is independent of the system state $\bm{x}$, it does not contribute
to the covariance. Equation$~\eqref{eq:supp-universal-response}$
then reduces to the standard equilibrium response relation
\begin{equation}
R_{A}(\lambda)=\left\langle \partial_{\lambda}A\right\rangle _{\lambda}-\beta\Cov_{\lambda}\!\left(A,\partial_{\lambda}H_{\mathrm{o}}\right).\label{eq:supp-equilibrium-response}
\end{equation}

\subsection{Exact stationary transport}

Let the control parameter vary according to a differentiable protocol
$\lambda=\lambda(t)$ for $0\leq t\leq\tau$. When the parameter changes
at a finite rate, the bare dynamics generally cannot adjust immediately
to the changing stationary state. As a result, the actual density
$p(\bm{x},t)$ may lag behind the instantaneous stationary density
$\rho_{\mathrm{st}}(\bm{x},\lambda(t))$.

Following the shortcut-to-isothermality method of \citep{LiQuanTu2017}
and the flow-field method of \citep{PatraJarzynski2017}, we introduce
an auxiliary control $\bm{f}(\bm{x},t)$. With the sign convention
used in the main text, the controlled dynamics are
\begin{equation}
\mathrm{d}\bm{X}_{t}=\left[\bm{b}_{\mathrm{o}}(\bm{X}_{t},\lambda(t))-\bm{f}(\bm{X}_{t},t)\right]\mathrm{d}t+\sqrt{2D}\,\mathrm{d}\bm{W}_{t},
\end{equation}
and the corresponding Fokker-Planck equation is
\begin{equation}
\partial_{t}p=\mathcal{L}_{\lambda(t)}^{\dagger}p+\nabla\cdot\left[\bm{f}(\bm{x},t)p(\bm{x},t)\right].
\end{equation}

Our goal is to keep the density equal to the instantaneous stationary
density throughout the protocol:
\begin{equation}
p(\bm{x},t)=\rho_{\mathrm{st}}(\bm{x},\lambda(t)).
\end{equation}
Substituting this expression into the Fokker-Planck equation and using
$\mathcal{L}_{\lambda}^{\dagger}\rho_{\mathrm{st}}(\bm{x},\lambda)=0$,
we obtain
\begin{equation}
\nabla\cdot\left[\rho_{\mathrm{st}}(\bm{x},\lambda(t))\bm{f}(\bm{x},t)\right]=\dot{\lambda}(t)\,\partial_{\lambda}\rho_{\mathrm{st}}(\bm{x},\lambda(t)).\label{eq:supp-control-condition}
\end{equation}
This equation states how the auxiliary control must move probability
density as the parameter changes.

Because the stationary density is normalized,
\begin{equation}
\int_{\Omega}\rho_{\mathrm{st}}(\bm{x},\lambda)\,\mathrm{d}^{n}x=1,
\end{equation}
we have
\begin{equation}
\int_{\Omega}\partial_{\lambda}\rho_{\mathrm{st}}(\bm{x},\lambda)\,\mathrm{d}^{n}x=\partial_{\lambda}\int_{\Omega}\rho_{\mathrm{st}}(\bm{x},\lambda)\,\mathrm{d}^{n}x=0.
\end{equation}
Thus, changing $\lambda$ redistributes probability without changing
the total probability.

Under periodic or no-flux boundary conditions, or when the density
decays sufficiently fast at infinity, we introduce a sufficiently
regular vector field $\bm{u}(\bm{x},\lambda)$ satisfying
\begin{equation}
\partial_{\lambda}\rho_{\mathrm{st}}(\bm{x},\lambda)=\nabla\cdot\left[\rho_{\mathrm{st}}(\bm{x},\lambda)\bm{u}(\bm{x},\lambda)\right].\label{eq:supp-transport}
\end{equation}
We refer to $\bm{u}$ as the escort field. It describes how probability
must be moved when the parameter is changed by a small amount.

Multiplying Eq.$~\eqref{eq:supp-transport}$ by $\dot{\lambda}(t)$
and comparing it with Eq.$~\eqref{eq:supp-control-condition}$ gives
\begin{equation}
\nabla\cdot\left\{ \rho_{\mathrm{st}}\left[\bm{f}-\dot{\lambda}\bm{u}\right]\right\} =0,
\end{equation}
where all quantities are evaluated at $(\bm{x},\lambda(t))$. Therefore,
the difference $\bm{f}-\dot{\lambda}\bm{u}$ has zero divergence after
being weighted by $\rho_{\mathrm{st}}$. Such a term may change the
probability current, but it does not change the density evolution.
Since our aim is only to transport the density, we choose the simplest
control and omit this additional current. The auxiliary control then
becomes
\begin{equation}
\boxed{\bm{f}(\bm{X}_{t},t)=\dot{\lambda}(t)\,\bm{u}(\bm{X}_{t},\lambda(t))}.\label{eq:supp-factorized-control}
\end{equation}
Thus, the strength of the auxiliary control is proportional to the
speed of the protocol.

We next verify that this control produces exact stationary tracking.
We assume that $\bm{u}$ and the boundary conditions are regular enough
for the controlled Fokker-Planck equation to have a unique normalized
solution. Suppose that the initial density is
\begin{equation}
p(\bm{x},0)=\rho_{\mathrm{st}}(\bm{x},\lambda(0)).
\end{equation}
Using Eqs.$~\eqref{eq:supp-transport}$ and $\eqref{eq:supp-factorized-control}$,
we find
\begin{align}
\partial_{t}\rho_{\mathrm{st}}(\bm{x},\lambda(t)) & =\dot{\lambda}(t)\,\partial_{\lambda}\rho_{\mathrm{st}}(\bm{x},\lambda(t))\\
 & =\dot{\lambda}(t)\,\nabla\cdot\left[\rho_{\mathrm{st}}(\bm{x},\lambda(t))\bm{u}(\bm{x},\lambda(t))\right]\\
 & =\nabla\cdot\left[\rho_{\mathrm{st}}(\bm{x},\lambda(t))\bm{f}(\bm{x},t)\right].
\end{align}
Together with $\mathcal{L}_{\lambda(t)}^{\dagger}\rho_{\mathrm{st}}=0$,
this shows that $\rho_{\mathrm{st}}(\bm{x},\lambda(t))$ satisfies
the controlled Fokker-Planck equation. Because the solution is unique,
the actual density must be
\begin{equation}
p(\bm{x},t)=\rho_{\mathrm{st}}(\bm{x},\lambda(t)),\qquad0\leq t\leq\tau.
\end{equation}

The argument also shows the converse: if exact stationary tracking
is achieved, the control must satisfy Eq.$~\eqref{eq:supp-control-condition}$.
After removing any additional divergence-free current, this condition
reduces to Eq.$~\eqref{eq:supp-transport}$. Therefore, Eq.$~\eqref{eq:supp-transport}$
is both necessary and sufficient for exact stationary transport under
the simplest choice of auxiliary control.

Finally, if $\dot{\lambda}(t)$ vanishes at isolated times, the factorized
control remains well defined as long as $\bm{u}$ is regular. In particular,
choosing $\dot{\lambda}(0)=\dot{\lambda}(\tau)=0$ makes the auxiliary
control vanish at the beginning and end of the protocol.

\subsection{Effective reversible dynamics}

We now consider the case in which the stationary density is not available
in closed form. For the bare dynamics introduced in Sec.$~$I A, the
probability current at fixed $\lambda$ is
\begin{equation}
\bm{J}_{0}[p]=\bm{b}_{\mathrm{o}}(\bm{x},\lambda)p-D\nabla p.\label{eq:supp-bare-current}
\end{equation}
At stationarity,
\begin{equation}
\nabla\cdot\bm{J}_{\mathrm{st}}=0,\qquad\bm{J}_{\mathrm{st}}=\bm{b}_{\mathrm{o}}\rho_{\mathrm{st}}-D\nabla\rho_{\mathrm{st}}.\label{eq:supp-stationary-current}
\end{equation}
Detailed balance implies $\bm{J}_{\mathrm{st}}=0$, whereas a nonequilibrium
steady state may sustain a nonzero divergence-free current \citep{Seifert2012}.

The spatial dependence of the stationary density is characterized
by the score
\begin{equation}
\bm{s}(\bm{x},\lambda)\equiv\nabla\ln\rho_{\mathrm{st}}(\bm{x},\lambda).\label{eq:supp-score}
\end{equation}
Using this definition, Eq.$~\eqref{eq:supp-stationary-current}$ gives
\begin{equation}
\bm{b}_{\mathrm{o}}(\bm{x},\lambda)=D\bm{s}(\bm{x},\lambda)+\frac{\bm{J}_{\mathrm{st}}(\bm{x},\lambda)}{\rho_{\mathrm{st}}(\bm{x},\lambda)}.\label{eq:supp-drift-decomposition}
\end{equation}
The stationary density determines the score term $D\bm{s}$ but not
the stationary-current contribution. Therefore, the bare drift $\bm{b}_{\mathrm{o}}$
cannot in general be reconstructed from the stationary density alone.

We define the effective reversible drift
\begin{equation}
\bm{b}_{\mathrm{eff}}(\bm{x},\lambda)\equiv D\bm{s}(\bm{x},\lambda).\label{eq:supp-effective-drift}
\end{equation}
At the target stationary density, the probability current generated
by this drift is
\begin{align}
\bm{J}_{\mathrm{eff}}[\rho_{\mathrm{st}}] & =\bm{b}_{\mathrm{eff}}\rho_{\mathrm{st}}-D\nabla\rho_{\mathrm{st}}\nonumber \\
 & =D\rho_{\mathrm{st}}\nabla\ln\rho_{\mathrm{st}}-D\nabla\rho_{\mathrm{st}}=0.\label{eq:supp-effective-current}
\end{align}
Hence the effective dynamics
\begin{equation}
\dd\bm{X}_{t}=\bm{b}_{\mathrm{eff}}(\bm{X}_{t},\lambda)\,\dd t+\sqrt{2D}\,\dd\bm{W}_{t}\label{eq:supp-effective-sde}
\end{equation}
have $\rho_{\mathrm{st}}(\bm{x},\lambda)$ as a stationary density.
The bare and effective dynamics therefore share the same stationary
density, but they generally have different stationary currents, entropy
production rates, transition pathways, and time-correlation functions.
At this stage, only equality of their stationary densities has been
established. 

Because the stationary-transport construction requires only a reference
drift that preserves $\rho_{\mathrm{st}}$ at fixed $\lambda$, the
effective drift may be used in place of $\bm{b}_{\mathrm{o}}$ during
a parameter sweep. The controlled dynamics then become
\begin{equation}
\dd\bm{X}_{t}=\left[D\bm{s}(\bm{X}_{t},\lambda(t))-\dot{\lambda}(t)\bm{u}(\bm{X}_{t},\lambda(t))\right]\dd t+\sqrt{2D}\,\dd\bm{W}_{t}.\label{eq:supp-effective-controlled-sde}
\end{equation}
If the score and escort field are exact, the transport equation and
prescribed initial condition, together with the well-posedness assumptions
above, ensure $p(\bm{x},t)=\rho_{\mathrm{st}}(\bm{x},\lambda(t))$
throughout the protocol. The stationary density can therefore be transported
without reconstructing the original bare drift.

At equilibrium, $\rho_{\mathrm{st}}\propto\exp[-\beta H_{\mathrm{o}}(\bm{x},\lambda)]$
and $\bm{s}=-\beta\nabla H_{\mathrm{o}}$. In the unit-mobility convention,
$D=\beta^{-1}$ and hence $\bm{b}_{\mathrm{eff}}=-\nabla H_{\mathrm{o}}$,
recovering the usual overdamped equilibrium drift.

\subsection{Operational response formulas}

We now derive an operational form of the stationary-response identity
for a density that cannot be evaluated explicitly. Dividing Eq.$~\eqref{eq:supp-transport}$
by the strictly positive stationary density gives
\begin{align}
\partial_{\lambda}\ln\rho_{\mathrm{st}} & =\frac{1}{\rho_{\mathrm{st}}}\nabla\cdot(\rho_{\mathrm{st}}\bm{u})\nonumber \\
 & =\nabla\cdot\bm{u}+\bm{u}\cdot\nabla\ln\rho_{\mathrm{st}}\nonumber \\
 & =\nabla\cdot\bm{u}+\bm{u}\cdot\bm{s}.\label{eq:supp-log-density-derivative}
\end{align}
Define
\begin{equation}
G(\bm{x},\lambda)\equiv\partial_{\lambda}\ln\rho_{\mathrm{st}}(\bm{x},\lambda)=\nabla\cdot\bm{u}(\bm{x},\lambda)+\bm{u}(\bm{x},\lambda)\cdot\bm{s}(\bm{x},\lambda).\label{eq:supp-G-definition}
\end{equation}
Normalization implies
\begin{equation}
\langle G\rangle_{\lambda}=\int_{\Omega}\rho_{\mathrm{st}}G\,\dd^{n}x=\int_{\Omega}\partial_{\lambda}\rho_{\mathrm{st}}\,\dd^{n}x=0.\label{eq:supp-G-zero-mean}
\end{equation}
Since $\partial_{\lambda}\Phi=-G$, substitution into Eq.$~\eqref{eq:supp-universal-response}$
yields 
\begin{equation}
\boxed{R_{A}(\lambda)=\left\langle \partial_{\lambda}A\right\rangle _{\lambda}+\Cov_{\lambda}(A,G)}.\label{eq:supp-operational-response-formula}
\end{equation}
For an observable with no explicit dependence on $\lambda$, the first
term on the right-hand side vanishes.

Equation$~\eqref{eq:supp-operational-response-formula}$ also admits
a local form that avoids explicit evaluation of either the score or
the divergence of the escort field. Since $\langle G\rangle_{\lambda}=0$,
its covariance term can be written as
\begin{align}
\Cov_{\lambda}(A,G) & =\langle AG\rangle_{\lambda}\nonumber \\
 & =\int_{\Omega}A(\bm{x},\lambda)\nabla\cdot\left[\rho_{\mathrm{st}}(\bm{x},\lambda)\bm{u}(\bm{x},\lambda)\right]\dd^{n}x,\label{eq:supp-covariance-divergence-form}
\end{align}
where Eq.$~\eqref{eq:supp-G-definition}$ was used in the second line.
Applying the divergence theorem and integration by parts gives
\begin{align}
\Cov_{\lambda}(A,G) & =\int_{\partial\Omega}A\rho_{\mathrm{st}}\bm{u}\cdot\bm{n}\,\dd S-\int_{\Omega}\rho_{\mathrm{st}}\bm{u}\cdot\nabla A\,\dd^{n}x\nonumber \\
 & =\int_{\partial\Omega}A\rho_{\mathrm{st}}\bm{u}\cdot\bm{n}\,\dd S-\left\langle \bm{u}\cdot\nabla A\right\rangle _{\lambda}.\label{eq:supp-response-integration-by-parts}
\end{align}
If the boundary conditions are such that the surface term in this
equation vanishes, the response reduces to the local form
\begin{equation}
\boxed{R_{A}(\lambda)=\left\langle \partial_{\lambda}A\right\rangle _{\lambda}-\left\langle \bm{u}\cdot\nabla A\right\rangle _{\lambda}}.\label{eq:supp-local-response-formula}
\end{equation}
This expression is equivalent to Eq.$~\eqref{eq:supp-operational-response-formula}$
for an exact escort field satisfying the transport equation. It is
useful computationally because it requires only the escort field and
the spatial gradient of the observable. 

For a fixed instantaneous state observable $A(\bm{x},\lambda)$ defined
independently of the dynamics, Eq.$~\eqref{eq:supp-operational-response-formula}$
depends only on the parameterized stationary density and not on the
stationary probability current. The physical and effective dynamics
therefore give the same static stationary response for the same state
observable. This conclusion does not apply to observables that explicitly
involve the physical drift, probability current, or entropy production,
nor does it apply to time-correlation or trajectory-dependent response
functions. 

In the learned implementation for interacting many-body systems, we
parameterize approximations to the score and the escort field by graph
neural networks (GNNs) $\bm{s}_{\phi}$ and $\bm{u}_{\theta}$, respectively:
\begin{equation}
\bm{s}_{\phi}\approx\bm{s},\qquad\bm{u}_{\theta}\approx\bm{u}.\label{eq:supp-learned-fields}
\end{equation}
We define the surrogate
\begin{equation}
\widehat{G}_{\theta,\phi}(\bm{x},\lambda)\equiv\nabla\cdot\bm{u}_{\theta}(\bm{x},\lambda)+\bm{u}_{\theta}(\bm{x},\lambda)\cdot\bm{s}_{\phi}(\bm{x},\lambda).\label{eq:supp-learned-G}
\end{equation}
This surrogate equals $G$ only in the limit of exact score and escort
fields. Using samples intended to represent the stationary ensemble,
the corresponding field-based response estimator is
\begin{equation}
\widehat{R}_{A}(\lambda)=\left\langle \partial_{\lambda}A\right\rangle _{\lambda}+\Cov_{\lambda}\!\left(A,\widehat{G}_{\theta,\phi}\right).\label{eq:supp-learned-response}
\end{equation}
The integration-by-parts form gives the alternative local estimator
\begin{equation}
\widehat{R}_{A}^{\mathrm{loc}}(\lambda)=\left\langle \partial_{\lambda}A\right\rangle _{\lambda}-\left\langle \bm{u}_{\theta}\cdot\nabla A\right\rangle _{\lambda},\label{eq:supp-learned-local-response}
\end{equation}
which does not require the learned score or the divergence of $\bm{u}_{\theta}$.
In general, $\langle\widehat{G}_{\theta,\phi}\rangle_{\lambda}$ need
not vanish, so the covariance in Eq.$~\eqref{eq:supp-learned-response}$
must remain centered. With approximate learned fields, residual score
error, transport error, and distributional lag make Eq.$~\eqref{eq:supp-learned-response}$
an estimator rather than an exact identity. The same learned fields
can nevertheless be reused to estimate the stationary responses of
different state observables.

The complete STPS workflow is summarized in the following algorithm.
The equilibrium and NESS learning procedures are detailed in Secs.$~$II
and III, respectively.

\begin{algorithm}[t]

\caption{Stationary-response estimation with shortcuts to parameter sweeps}

\label{alg:response-auxiliary-control}

\KwIn{
Stationary reference sample sets
$\{\mathcal{S}_{\lambda_j}\}_{j=1}^{N_\lambda}$, including a set at
$\lambda(0)$;
a scalar state observable $A(\boldsymbol{x},\lambda)$;
a differentiable protocol $\lambda:[0,\tau]\to I$;
the diffusion coefficient $D$;
and the number $M$ of controlled replicas.
For an equilibrium system, also provide
$H_{\mathrm{o}}(\boldsymbol{x},\lambda)$ and $\beta$.
}

\KwOut{
The estimated response curve $\widehat{R}_A(\lambda)$ over the swept
parameter interval.
}

\textbf{Construct the stationary score.}
For an equilibrium system, use
\[
\boldsymbol{s}(\boldsymbol{x},\lambda)
=
-\beta\nabla_{\boldsymbol{x}}
H_{\mathrm{o}}(\boldsymbol{x},\lambda).
\]

For a NESS with unknown stationary density, learn
\[
\boldsymbol{s}_{\phi}(\boldsymbol{x},\lambda)
\simeq
\nabla_{\boldsymbol{x}}
\ln\rho_{\mathrm{st}}(\boldsymbol{x},\lambda).
\]
\;

\textbf{Obtain the escort field.}
For an equilibrium system, construct $\boldsymbol{u}$ analytically
when available. Otherwise---and for the NESS branch considered
here---learn $\boldsymbol{u}_{\theta}\simeq\boldsymbol{u}$ from the
stationary reference samples. The exact escort field satisfies
\[
\partial_{\lambda}\rho_{\mathrm{st}}(\boldsymbol{x},\lambda)
=
\nabla_{\boldsymbol{x}}\cdot
\left[
\rho_{\mathrm{st}}(\boldsymbol{x},\lambda)
\boldsymbol{u}(\boldsymbol{x},\lambda)
\right].
\]

Using the auxiliary-control notation of the main text, set
\[
\boldsymbol{f}(\boldsymbol{x},t)
=
\begin{cases}
\dot{\lambda}(t)
\boldsymbol{u}\bigl(\boldsymbol{x},\lambda(t)\bigr),
& \text{exact construction},
\\[1mm]
\dot{\lambda}(t)
\boldsymbol{u}_{\theta}
\bigl(\boldsymbol{x},\lambda(t)\bigr),
& \text{learned construction}.
\end{cases}
\]
\;

\textbf{Run a controlled ensemble sweep.}
Draw $M$ independent initial conditions from the stationary reference
set at $\lambda(0)$, corresponding to
\[
\boldsymbol{x}^{(m)}_0
\overset{\mathrm{i.i.d.}}{\sim}
\rho_{\mathrm{st}}\bigl(\,\cdot\,,\lambda(0)\bigr),
\qquad
m=1,\ldots,M.
\]

For each $m$, evolve an independent copy of $\boldsymbol{X}_t$ with
$\boldsymbol{X}_0=\boldsymbol{x}^{(m)}_0$.
For an equilibrium system, use the bare drift
$b_{\mathrm{o}}=-D\beta\nabla_{\boldsymbol{x}}H_{\mathrm{o}}$;
for a NESS with unknown stationary density, use
$D\boldsymbol{s}_{\phi}$ as the learned approximation to the effective
reversible drift
$\boldsymbol{b}_{\mathrm{eff}}=D\boldsymbol{s}$.
Simulate
\[
\mathrm{d}\boldsymbol{X}_t
=
\begin{cases}
\Bigl[
-D\beta\nabla_{\boldsymbol{x}}H_{\mathrm{o}}
\bigl(\boldsymbol{X}_t,\lambda(t)\bigr)
-\boldsymbol{f}(\boldsymbol{X}_t,t)
\Bigr]\mathrm{d}t
+\sqrt{2D}\,\mathrm{d}\boldsymbol{W}_t,
& \text{equilibrium},
\\[2mm]
\Bigl[
D\boldsymbol{s}_{\phi}
\bigl(\boldsymbol{X}_t,\lambda(t)\bigr)
-\boldsymbol{f}(\boldsymbol{X}_t,t)
\Bigr]\mathrm{d}t
+\sqrt{2D}\,\mathrm{d}\boldsymbol{W}_t,
& \text{NESS with unknown }\rho_{\mathrm{st}}.
\end{cases}
\]
\;

\textbf{Evaluate the response.}
At each recorded value $\lambda=\lambda(t)$, evaluate ensemble averages
and covariances over the $M$ controlled replicas.
For the NESS branch, first compute
\[
\widehat{G}_{\theta,\phi}(\boldsymbol{x},\lambda)
=
\nabla_{\boldsymbol{x}}\cdot
\boldsymbol{u}_{\theta}(\boldsymbol{x},\lambda)
+
\boldsymbol{u}_{\theta}(\boldsymbol{x},\lambda)
\cdot
\boldsymbol{s}_{\phi}(\boldsymbol{x},\lambda).
\]

Then return
\[
\widehat{R}_A(\lambda)
=
\begin{cases}
\bigl\langle\partial_{\lambda}A\bigr\rangle_{\lambda}
-\beta\operatorname{Cov}_{\lambda}\!\left(
A,\partial_{\lambda}H_{\mathrm{o}}
\right),
& \text{equilibrium},
\\[2mm]
\bigl\langle\partial_{\lambda}A\bigr\rangle_{\lambda}
+\operatorname{Cov}_{\lambda}\!\left(
A,\widehat{G}_{\theta,\phi}
\right),
& \text{NESS with unknown }\rho_{\mathrm{st}}.
\end{cases}
\]
\;

\end{algorithm}

\section{Learning the escort field from equilibrium samples}

We consider a system with potential $U_{\mathrm{o}}(\bm{x};\lambda)$
and seek an auxiliary control that keeps the driven system close to
the instantaneous equilibrium density. The GNN represents a rate-independent
field $\bm{u}_{\theta}(\bm{x},\lambda(t))$. For a protocol $\lambda(t)$,
the auxiliary control and the controlled drift are defined as $\bm{f}_{\mathrm{aux}}(\bm{x},t)=\dot{\lambda}(t)\,\bm{u}_{\theta}(\bm{x},\lambda(t))$
and $\bm{b}_{\mathrm{ctrl}}=-\bm{\nabla}_{\bm{x}}U_{\mathrm{o}}-\bm{f}_{\mathrm{aux}}.$
The protocol speed $\dot{\lambda}$ is multiplied outside the network
and is not used as a network input. For a system of $10$ particles
in two dimensions, the auxiliary field is represented by a fully connected
GNN that is equivariant under particle permutations. It contains three
message-passing blocks \citep{Gilmer2017} with hidden dimension 64,
SiLU activations, and layer normalization. Its output has the form
\begin{equation}
\bm{u}_{\theta,i}=a_{i}\bm{x}_{i}+\sum_{j\ne i}s_{ij}(\bm{x}_{i}-\bm{x}_{j}),
\end{equation}
where the scalar coefficients $a_{i}$ and $s_{ij}$ are produced
by the GNN. This construction preserves the symmetries of the particle
system \citep{Satorras2021,Batzner2022}. The final network layers
are initialized to zero, so that training starts from an almost vanishing
auxiliary control. We next describe how this field is learned from
equilibrium data.

At each value of $\lambda$, equilibrium configurations provide the
reference distribution. Starting from these configurations, we integrate
the controlled dynamics differentiably and compare the resulting ensemble
with the instantaneous equilibrium ensemble. The training loss is
\begin{equation}
\mathcal{L}=2\mathcal{L}_{\mathrm{mom}}+2\mathcal{L}_{\mathrm{energy}}+10^{-4}\left\langle \lVert\bm{u}_{\theta}\rVert^{2}\right\rangle .
\end{equation}
Here, $\mathcal{L}_{\mathrm{mom}}$ matches permutation- and rotation-invariant
second moments, while $\mathcal{L}_{\mathrm{energy}}$ matches equilibrium
energy statistics. No target auxiliary field is provided. Instead,
the network is trained by matching the statistics of the controlled
ensemble to those of the equilibrium reference ensemble. 

The equilibrium reference bank covers $\lambda\in[1.60,4.40]$ with
a spacing of $0.025$, giving $113$ parameter values. We generated
$24,576$ equilibrium configurations at each parameter value, for
a total of $2,777,088$ reference samples. This wider interval places
the target range $\lambda\in[2,4]$ strictly inside the training domain. 

Training used a linear protocol from $1.6$ to $4.4$ with duration
$\tau=2$. Each optimization update used a batch size of 8192 and
a differentiable rollout of duration 0.1, implemented with the Euler--Maruyama
scheme \citep{KloedenPlaten1992} and a time step of $10^{-3}$. The
network was trained from scratch using Adam \citep{KingmaBa2015},
with an initial learning rate of $3\times10^{-4}$ followed by cosine
decay and gradient clipping at a global norm of $5$. We define one
training epoch as $100$ optimization updates. Training was run for
$60$ epochs, corresponding to $6000$ updates in total. The checkpoint
obtained after the final update was used for all reported results.
Final response curves were evaluated using $10^{5}$ independent trajectories
with random seeds not used during training.

\section{Learning the score and escort field from NESS samples}

The score and escort networks were optimized in four separate stages.
We first trained an initial score model and held it fixed while obtaining
an initial escort field. The score was then refined independently.
Finally, the refined score was frozen and used to refine the escort
field initialized in the first escort stage. The two networks were
never updated jointly. Network training used only stationary configurations
and their parameter values; the physical drift was used to generate
the stationary data but did not enter any loss. 

\subsection{Score learning}

For the nonequilibrium steady state (NESS), we approximate the score
by 
\begin{equation}
\bm{s}_{\phi}(\bm{x},\lambda)\simeq\nabla_{\bm{x}}\ln\rho_{\mathrm{st}}(\bm{x},\lambda).\label{eq:supp-ness-score-model}
\end{equation}
Each configuration of $N=8$ two-dimensional particles is represented
as a complete directed graph without self-edges. The node inputs are
$(\bm{x}_{i},\lVert\bm{x}_{i}\rVert^{2})$, the edge inputs are $\bigl(\bm{x}_{i}-\bm{x}_{j},r_{ij}^{2},\exp[-r_{ij}^{2}/(2\sigma^{2})]\bigr)$,
and $\lambda$ is supplied as a global input. The network contains
three message-passing blocks with hidden dimension 128, SiLU activations,
and layer normalization, and returns one two-dimensional vector per
particle. Shared node and edge maps make the output equivariant under
particle permutations. Inversion symmetry is imposed through $\bm{s}_{\phi}(-\bm{x},\lambda)=-\bm{s}_{\phi}(\bm{x},\lambda).$
The final vector decoder of the initial network is zero-initialized. 

The primary objective is the Hyvärinen implicit score-matching loss
\citep{Hyvarinen2005}, 
\begin{equation}
\mathcal{L}_{\mathrm{ISM}}=\mathbb{E}_{\rho_{\mathrm{st}}}\left[\lVert\bm{s}_{\phi}\rVert^{2}+2\nabla_{\bm{x}}\!\cdot\bm{s}_{\phi}\right].\label{eq:supp-ness-ism}
\end{equation}
Let $m=2N$, flatten $\bm{x}$ and $\bm{s}_{\phi}$ into $m$-component
vectors, and define $\bm{M}_{\phi}\equiv\mathbb{E}_{\rho_{\mathrm{st}}}\left[\bm{x}\bm{s}_{\phi}^{\mathsf{T}}\right].$
When $\bm{s}_{\phi}$ is the exact score and the boundary term vanishes,
the Stein identity gives $\bm{M}_{\phi}=-\bm{I}_{m}$ \citep{Liu2016};
this relation is encouraged during training by
\begin{equation}
\mathcal{L}_{\mathrm{trS}}=\left(\operatorname{tr}\bm{M}_{\phi}+m\right)^{2},\qquad\mathcal{L}_{\mathrm{matS}}=\frac{1}{m}\left\lVert \bm{M}_{\phi}+\bm{I}_{m}\right\rVert _{F}^{2}.\label{eq:supp-ness-stein-losses}
\end{equation}
The initial stage used
\begin{equation}
\mathcal{L}_{\phi}^{(1)}=\mathcal{L}_{\mathrm{ISM}}+0.05\mathcal{L}_{\mathrm{trS}}+0.10\mathcal{L}_{\mathrm{matS}}.\label{eq:supp-ness-score-loss-stage1}
\end{equation}
The divergence in Eq.$~\eqref{eq:supp-ness-ism}$ was estimated with
four independent Rademacher Hutchinson probes per update \citep{Hutchinson1990}. 

During score refinement, the divergence was evaluated exactly from
the score Jacobian. To enforce the integrability of the learned field,
define $\mathcal{A}[\bm{J}]=\bm{J}-\bm{J}^{\mathsf{T}}$ and add
\begin{align}
 & \mathcal{L}_{\mathrm{curl}}=\mathbb{E}\left[\frac{\lVert\mathcal{A}[\nabla_{\bm{x}}\bm{s}_{\phi}]\rVert_{F}^{2}}{m^{2}}\right],\nonumber \\
 & \mathcal{L}_{\lambda\mathrm{curl}}=\mathbb{E}\left[\frac{\lVert\mathcal{A}[\nabla_{\bm{x}}\partial_{\lambda}\bm{s}_{\phi}]\rVert_{F}^{2}}{\lVert\nabla_{\bm{x}}\partial_{\lambda}\bm{s}_{\phi}\rVert_{F}^{2}+\epsilon}\right].\label{eq:supp-ness-score-curl-losses}
\end{align}
The raw refinement weights of these terms were $0.20$ and $0.10$,
respectively; $\partial_{\lambda}\bm{s}_{\phi}$ in the second penalty
was evaluated by a centered difference with step $10^{-2}$. The Stein
and curl gradients were projected to remove components that opposed
the exact-ISM gradient and were capped at $0.12$, $0.10$, and $0.35$
times its norm for the Stein, $\mathcal{L}_{\mathrm{curl}}$, and
$\mathcal{L}_{\lambda\mathrm{curl}}$ contributions, respectively.
Thus, exact score matching remained the primary refinement signal. 

Stationary configurations were generated at $113$ equally spaced
values of $\lambda\in[1.60,4.40]$ with spacing $0.025$. At each
value, $32,768$ independent configurations were retained after $12,000$
Euler-Maruyama relaxation steps of size $10^{-3}$, giving $3,702,784$
configurations in total. Both score stages used this stationary bank.
The initial stage used Adam \citep{KingmaBa2015}, batch size $2048$,
an initial learning rate of $10^{-3}$ with cosine decay, gradient
clipping at norm $5$, and an exponential moving average (EMA) decay
of $0.999$. It comprised $150$ epochs with $100$ updates per epoch,
or $15,000$ updates. 

The refinement stage was initialized from the parameters produced
by the initial stage. Each exact-ISM update used an effective batch
of $768$ distributed over three values of $\lambda$; the Stein penalties
used $1536$ configurations, and the $\lambda$-derivative penalty
used $80$ configurations distributed over five values in $[2,4]$.
Adam was reinitialized at a learning rate of $3\times10^{-6}$, which
was reduced during refinement to a minimum of $5\times10^{-7}$. The
EMA decay was $0.99$, and gradients were clipped at norm $5$. Refinement
comprised $80$ epochs with $30$ updates per epoch, or $2400$ updates.
The complete score training therefore contained $17,400$ Adam updates.
After the last refinement step, the resulting score model was frozen
and used throughout the subsequent escort-field training and response
evaluation. 

\subsection{Escort-field learning}

The escort field is represented by a second GNN $\bm{u}_{\theta}(\bm{x},\lambda)$.
For a protocol $\lambda(t)$, the learned controlled dynamics are
\begin{equation}
\dd\bm{X}_{t}=\left[D\bm{s}_{\phi}(\bm{X}_{t},\lambda(t))-\dot{\lambda}(t)\bm{u}_{\theta}(\bm{X}_{t},\lambda(t))\right]\dd t+\sqrt{2D}\,\dd\bm{W}_{t}.\label{eq:supp-ness-controlled-sde}
\end{equation}
The simulations used $D=1$, while $D$ is retained in this equation
to display the general convention. The escort network uses the same
graph representation, hidden dimension, and three message-passing
blocks as the score network, but has independent parameters $\theta$.
It obeys $\bm{u}_{\theta}(-\bm{x},\lambda)=-\bm{u}_{\theta}(\bm{x},\lambda),$
and its initial vector decoder is zero-initialized. No target values
of the escort field are used. 

From the learned fields, define 
\begin{equation}
\widehat{G}_{\theta,\phi}(\bm{x},\lambda)=\nabla_{\bm{x}}\!\cdot\bm{u}_{\theta}+\bm{u}_{\theta}\!\cdot\bm{s}_{\phi}.\label{eq:supp-ness-learned-G}
\end{equation}
The divergence is computed from the vector output by exact automatic
differentiation and is not fitted to divergence labels. The corresponding
stationary-response estimator is 
\begin{equation}
\widehat{R}_{A}(\lambda)=\left\langle \partial_{\lambda}A\right\rangle _{\mathrm{st}}+\Cov_{\mathrm{st}}\left(A,\widehat{G}_{\theta,\phi}\right).\label{eq:supp-ness-response-estimator}
\end{equation}
Because the fields are learned approximations, this equation is an
estimator rather than an exact identity. 

The pointwise gradient-transport consistency loss is 
\begin{equation}
\mathcal{L}_{\mathrm{GT}}=\mathbb{E}_{\rho_{\mathrm{st}}}\left[\left\lVert \frac{\nabla_{\bm{x}}\widehat{G}_{\theta,\phi}-\partial_{\lambda}\bm{s}_{\phi}}{e_{\phi}}\right\rVert ^{2}\right],\qquad e_{\phi}=\left(\mathbb{E}\lVert\partial_{\lambda}\bm{s}_{\phi}\rVert^{2}\right)^{1/2}+0.1.\label{eq:supp-ness-gradient-transport-loss}
\end{equation}
Here $\partial_{\lambda}\bm{s}_{\phi}$ is obtained by a Jacobian-vector
product, while $\nabla_{\bm{x}}\widehat{G}_{\theta,\phi}$ and $\nabla\cdot\bm{u}_{\theta}$
are evaluated by automatic differentiation. This loss fixes $\widehat{G}_{\theta,\phi}$
only up to a $\lambda$-dependent constant. Such a constant cancels
from the covariance in Eq.$~\eqref{eq:supp-ness-response-estimator}$;
accordingly, no separate mini-batch penalty on the mean of $\widehat{G}_{\theta,\phi}$
was optimized. 

To constrain the remaining low-frequency components, we use the seven
coordinate observables 
\begin{equation}
\mathcal{B}=\left\{ \sum_{i}x_{i}^{2},\sum_{i}y_{i}^{2},\sum_{i}x_{i}y_{i},\sum_{i}x_{i}^{2}y_{i}^{2},\sum_{i}x_{i}^{4},\sum_{i}y_{i}^{4},\sum_{i<j}e^{-r_{ij}^{2}/(2\sigma^{2})}\right\} .\label{eq:supp-ness-lowfreq-observables}
\end{equation}
For each $B_{k}\in\mathcal{B}$, a sixth-order polynomial was fitted
to the bank estimate of $\langle B_{k}\rangle_{\mathrm{st}}$ over
$\lambda\in[1.60,4.40]$. Analytic differentiation of this fitted
mean curve provides the target $c_{k}(\lambda)=\partial_{\lambda}\langle B_{k}\rangle_{\mathrm{st}}$
for $\Cov(B_{k},\widehat{G}_{\theta,\phi})$. The resulting normalized
covariance loss is 
\begin{equation}
\mathcal{L}_{\mathrm{LF}}=\frac{1}{7}\sum_{k=1}^{7}\left[\frac{\widehat{\Cov}(B_{k},\widehat{G}_{\theta,\phi})-c_{k}(\lambda)}{a_{k}}\right]^{2},\label{eq:supp-ness-lowfreq-loss}
\end{equation}
where $a_{k}$ is the corresponding target scale. During refinement,
the sample covariance was computed using the unbiased denominator
$N_{b}-1$, where $N_{b}$ denotes the number of configurations in
the covariance batch. A conditioning channel formed from the known
linear combination of the $x^{2}$, $y^{2}$, and pair terms in the
potential was included with weight $2$; it introduced no escort labels
or additional response data. 

Short differentiable rollouts provide a complementary moment-matching
loss $\mathcal{L}_{\mathrm{MM}}$. It matches the one-particle and
distinct-particle second-moment tensors and the mean and variance
of the pair-kernel observable to their stationary-bank values at the
end of each segment. A penalty $\mathcal{L}_{\mathrm{reg}}=\langle\lVert\bm{u}_{\theta}\rVert^{2}\rangle$
controls the field magnitude. Thus, escort learning combines moment
matching, the gradient-transport loss, the low-frequency covariance
loss, and magnitude regularization, without reference escort fields. 

In the initial escort stage, the initial score model was held fixed.
Training used $28,672$ configurations per value of $\lambda$. Each
update used a rollout batch of $2048$ with three segments of duration
$0.1$ and $100$ Euler-Maruyama steps per segment. Both sweep directions
were sampled with protocol duration $\tau=2$. The gradient-transport
and low-frequency terms used $64$ configurations at each of three
values of $\lambda$. The outer transport weight was $0.20$, and
the relative low-frequency weight was $0.50$; the second-moment,
pair-feature, and field-magnitude weights were $2$, $1$, and $10^{-4}$.
Adam was initialized at $3\times10^{-4}$ with cosine decay, the EMA
decay was $0.995$, and gradients were clipped at norm $5$. This
stage comprised $80$ epochs with $80$ updates per epoch, or $6400$
updates. 

The initial escort parameters were refined using the final frozen
score and the full stationary bank. Each update used $64$ configurations
at each of four values of $\lambda$ for the gradient-transport term
and $1024$ configurations at one value for the low-frequency loss.
Every eight updates, a moment-preserving rollout used $1024$ configurations
over three values of $\lambda$, with three segments of duration $0.1$
and $50$ integration steps per segment. The low-frequency loss was
the primary gradient; projected transport and moment-matching gradients
were capped at $0.50$ and $0.10$ of its norm. The outer transport,
relative low-frequency, and magnitude weights were $0.08$, $0.75$,
and $2\times10^{-5}$. Adam was reinitialized at $10^{-5}$ and reduced
to a minimum of $10^{-6}$. The EMA decay was $0.99$, and gradients
were clipped at norm $5$. Refinement comprised $120$ epochs with
$30$ updates per epoch, or $3600$ updates. Escort training therefore
contained $10,000$ Adam updates in total. The checkpoint obtained
after the final update was used for all reported results. Final response
curves were evaluated using $10^{5}$ independent trajectories with
random seeds not used during training.

\section{Methods used for comparison}

In addition to shortcuts to parameter sweeps (STPS), we evaluated
the stationary response using independent central finite differences,
Malliavin-weight sampling (MWS), and a theoretical or high-precision
stationary reference. All reference calculations were performed independently
of STPS. For the observable $A=U_{\mathrm{o}}$, the compared quantity
was $R(\lambda)=\frac{\partial}{\partial\lambda}\langle A(\bm{x},\lambda)\rangle_{\mathrm{st}}.$ 

\subsection{Independent central finite differences}

At each reported value of $\lambda$, two independent stationary ensembles
were generated at $\lambda-\delta$ and $\lambda+\delta$. The response
and its sampling error were estimated as
\begin{align}
 & \widehat{R}_{\mathrm{FD}}(\lambda)=\frac{\langle A_{+}\rangle-\langle A_{-}\rangle}{2\delta},\label{eq:centered estimator}\\
 & \langle A_{\pm}\rangle=\frac{1}{M}\sum_{m=1}^{M}A(\bm{x}_{\pm}^{(m)},\lambda\pm\delta),\\
 & \operatorname{SEM}(\widehat{R}_{\mathrm{FD}})=\frac{\sqrt{\operatorname{SEM}\!\left(\langle A_{+}\rangle\right)^{2}+\operatorname{SEM}\!\left(\langle A_{-}\rangle\right)^{2}}}{2\delta}.
\end{align}
The two branches used independent initial states and Brownian increments;
no common random numbers were used. The choice of $\delta$ involves
the usual tradeoff between finite-difference truncation error and
the amplification of function-evaluation or sampling noise \citep{Gill1983,More2012}.
For the centered estimator in Eq.$~\eqref{eq:centered estimator}$,
a Taylor expansion about $\lambda$ gives a leading truncation error
of $O(\delta^{2})$.

For all three systems, response points were $\lambda=2,2.1,\ldots,4$,
with $\delta=0.05$, $\Delta t=10^{-3}$, and $M=10^{5}$ replicas
per branch. The harmonic-oscillator branches were initialized from
the exact stationary marginal of the Euler-Maruyama scheme and evolved
for two time units. The equilibrium and nonequilibrium interacting
systems were relaxed for $10$ and $12$ time units, respectively.
Error bars show one standard error.

\subsection{Malliavin-weight sampling}

At each fixed $\lambda$, the bare dynamics in Eq.$~\eqref{eq:supp-bare-sde}$
was first relaxed to stationarity. Following the Malliavin-weight
method introduced in \citep{WarrenAllen2012,WarrenAllen2014}, the
scalar weight $q$ was then reset to zero and accumulated over a finite
window according to
\begin{equation}
dq_{t}=\frac{1}{\sqrt{2D}}\,\partial_{\lambda}\bm{b}_{\mathrm{\mathrm{o}}}(\bm{X}_{t},\lambda)\mathbin{\cdot}d\bm{W}_{t}.
\end{equation}
In the Euler-Maruyama implementation, the weight used the prepoint
drift derivative and the same noise increment as the state update.
With $A_{\lambda}\equiv\partial_{\lambda}A$, the centered estimator
was
\begin{equation}
\widehat{R}_{\mathrm{MWS}}=\langle A_{\lambda}\rangle+\langle Aq\rangle-\langle A\rangle\langle q\rangle.
\end{equation}
Its one-standard-error bar was computed from
\begin{align}
 & \psi=\left(A_{\lambda}-\langle A_{\lambda}\rangle\right)+\left(Aq-\langle Aq\rangle\right)-\langle q\rangle\left(A-\langle A\rangle\right)-\langle A\rangle\left(q-\langle q\rangle\right),\\
 & \operatorname{SEM}(\widehat{R}_{\mathrm{MWS}})=\frac{\operatorname{std}(\psi)}{\sqrt{M}}.
\end{align}

All systems used the finite-difference parameter grid, $M=10^{5}$
trajectories, a $10$-time-unit relaxation, a weight-accumulation
window $\tau_{q}=2$, and $\Delta t=10^{-3}$. For the harmonic oscillator,
$\partial_{\lambda}b=1$ and $A_{\lambda}=-x$. For the equilibrium
system, $\partial_{\lambda}b_{i}=-r_{i}$ and $A_{\lambda}=\frac{1}{2}\sum_{i}\|r_{i}\|^{2}$.
For the nonequilibrium system, $\partial_{\lambda}b_{i}=(0,-y_{i})$
and $A_{\lambda}=\frac{1}{2}\sum_{i}y_{i}^{2}$. The reported SEM
measures sampling uncertainty only and does not include the systematic
truncation error from finite $\tau_{q}$. Neither MWS nor finite differences
used the learned score, the auxiliary control, or STPS trajectories.

\subsection{Theoretical and stationary references}

For the harmonic oscillator, $U_{\mathrm{o}}(x,\lambda)=kx^{2}/2-\lambda x$.
Its stationary mean energy and response are
\[
\langle U_{\mathrm{o}}\rangle_{\mathrm{st}}=\frac{1}{2\beta}-\frac{\lambda^{2}}{2k},\qquad R(\lambda)=-\frac{\lambda}{k}.
\]
Thus, for $k=2$, the analytical reference is $-\lambda/2$.

For the equilibrium interacting system, the canonical identity \citep{Zhu1993}
gives
\begin{equation}
\begin{aligned} & R_{\mathrm{eq}}(\lambda)=\langle O_{2}\rangle-\beta\operatorname{Cov}(U_{\mathrm{o}},O_{2}),\\
 & O_{2}=\partial_{\lambda}U_{\mathrm{o}}=\frac{1}{2}\sum_{i}\lVert\bm{r}_{i}\rVert^{2}.
\end{aligned}
\end{equation}
This expression was evaluated on a $0.05$ parameter grid using eight
post-relaxation snapshots of $10^{5}$ replicas per point; the standard
error was obtained by delta-method propagation of the sampled moments
\citep{Oehlert1992}.

For the nonequilibrium interacting system, no closed-form stationary
density is known. At each of $113$ banked parameter values, the mean
energy was evaluated from $32,768$ stationary configurations. A sixth-degree
polynomial $p_{6}$ was fitted to these means, and the reference was
defined by
\begin{equation}
R_{\mathrm{NESS}}^{\mathrm{ref}}(\lambda)=p_{6}'(\lambda).
\end{equation}
The curve labelled ‘Theory’ is therefore a high-precision stationary-data
reference, not a closed-form prediction.

\bibliographystyle{apsrev4-1}
\bibliography{refs_supp}